\documentclass[ twocolumn,08 pt, amsmath,amssymb, aps,fleqn]{revtex4-1}
\usepackage[hidelinks]{hyperref}
\usepackage[english]{babel}
\usepackage{mathtools}
\usepackage{tabularx}
\usepackage{multirow}
\usepackage{comment}
\usepackage{graphicx}
\usepackage{color}
\usepackage{soul}
\begin{document}
\preprint{APS/123-QED}
\title{Transport properties of a Luttinger liquid with a cluster of impurities}
\author{Joy Prakash Das} \author{Girish S. Setlur}\email{gsetlur@iitg.ernet.in}
\affiliation{Department of Physics \\ Indian Institute of Technology  Guwahati \\ Guwahati, Assam 781039, India}
\begin{abstract}
In this work, the correlation functions of a Luttinger liquid with a cluster of impurities around an origin obtained using the Non chiral bosonization technique (NCBT) are used to study two important physical phenomena, viz., conductance and resonant tunneling. The latter is studied when the cluster consists of two impurities separated by a distance (measured in units of the Fermi wavelength). Conductance is studied both in the Kubo formalism, which relates it to current-current correlations (four-point functions), as well as the outcome of a tunneling phenomena (two-point functions). In both the cases, closed analytical expressions for conductance are calculated and a number of interesting physical observations are discussed, besides presenting a favorable comparison with the existing literature.    
\end{abstract}

\maketitle
\section{Introduction}

One of the most important physical phenomena studied in condensed matter systems is the transport of electrons, especially when they are restricted to move in one dimension. This is because of the unique nature of the inter-particle interactions in one dimension which leads to interesting physics which is substantially different from that of the higher dimensions where interactions are tackled conveniently using the Fermi liquid theory. Secondly the emergence of advanced technologies has made the realization of one dimensional systems possible that have unusual properties and hold a promising future - carbon nanotubes \cite{bockrath1999luttinger}, semiconducting quantum wire \cite{auslaender2000experimental, yacoby1997magneto} and so on. The suitable alternative to the Fermi liquid theory to capture the many body physics of such 1D systems is the Luttinger liquid theory \cite{haldane1981luttinger} which has served as the paradigm for one dimensional systems and is based on linearization of the dispersion relations of the constituent particles near the Fermi level. 

Most of the physical phenomena of such systems can be systematically studied provided one has analytical forms of the correlation functions - to obtain these is the stated goal in quantum many body physics. In one dimension, this goal is achieved using bosonization methods where a fermion field operator is expressed as the exponential of a bosonic field \cite{von1998bosonization}.  This operator approach to bosonization, which goes under the name g-ology \cite{giamarchi2004quantum}, can be used successfully to compute the N-point Green functions of a clean Luttinger liquid. But the Fermi-Bose correspondence used in the g-ology methods is insufficient to tackle impurities and to circumvent this, other techniques like renormalization group (RG) methods are mandatory \cite{matveev1993tunneling}.

A novel technique by the name of `Non chiral bosonization technique' has been developed that uses a basis different from the plane wave basis to deal strongly inhomogeneous Luttinger liquid, without adhering to RG methods \cite{das2018quantum}.  NCBT can extract the most singular part of the correlation functions of a Luttinger liquid with arbitrary strength of the external impurities as well as that of mutual interactions between the particles. It has also been applied successfully to study the one step fermionic ladder (two 1D wires placed parallel and close to each other with hopping between a pair of opposing points) \cite{das2017one} and slowly moving heavy impurities in a Luttinger liquid \cite{das2018ponderous}. The Green functions enables one to predict different physical phenomena occurring in the system such as Friedel oscillations \cite{Egger1995friedel1}, conductance \cite{fendley1995exact, fendley1995exact2}, Kondo effect \cite{furusaki1994kondo, schiller1995exact}, resonant tunneling \cite{kane1992resonant, furusaki1993resonant}, etc. 

In the seminal work by Kane and Fisher \cite{kane1992transport}, it has been shown how impurities can bring drastic effects to the conductance of the particles which can be as severe as `cutting the chain' by even a small scatterer. Since then the study of transport phenomena in a Luttinger liquid with impurities has interested a number of researchers \cite{giamarchi1992conductivity, ogata1994collapse, safi1997conductance, ponomarenko1995renormalization}. The conductance of a narrow quantum wire with non-interacting electrons moving ballistically is given by $e^2/h$. This conductance is renormalized for a Luttinger liquid and is given by  $g e^2$/h, where g is the Luttinger liquid parameter which depends on the mutual interaction strength of the particles \cite{kane1992transport, apel1982combined, ogata1994collapse}. But no renormalization of the universal conductance is required if the electrons have a free behavior in the source and drain reservoirs \cite{ponomarenko1995renormalization, maslov1995landauer}. Matveev et al. used a simple renormalization group method to calculate the conductance of a weakly interacting electron gas in presence of a single scatterer \cite{matveev1993tunneling}. Ogata and Anderson \cite{ogata1993transport} used Green's functions to study conductivity of a Luttinger liquid and showed that if the spin-charge separation is taken into account, the resistivity has a linear temperature dependence. Besides conductance, resonant tunneling is yet another important phenomena studied in Luttinger liquid with double barriers \cite{kane1992resonant, kane1992transmission, furusaki1993resonant, moon1993resonant}. Kane and Fisher studied resonant tunneling in a single channel interacting electron gas through a double barrier and found that the width of the resonance vanishes, as a power of temperature, in the zero-temperature limit \cite {kane1992resonant, kane1992transmission}. Furusaki and Nagaosa studied the same for spinless fermions and calculated the conductance as a function of temperature and gate voltage \cite{furusaki1993resonant}. In another work, Furusaki studied resonant tunneling in a quantum dot weakly coupled to Luttinger liquids \cite{furusaki1998resonant} and a few years later, this model was supported by experimental evidences \cite{auslaender2000experimental}.

In this work, the conductance of a Luttinger liquid in presence of a cluster of impurity is calculated both in the Kubo formalism as well as the outcome of a tunneling experiment using the correlation functions obtained using NCBT. All the necessary limiting cases like Launderer's formula, conductance of a clean Luttinger liquid, half-line, etc. are all obtained. From the tunneling conductance the well known concepts of `cutting the chain' and `healing the chain' are elucidated. The condition of resonant tunneling for a double impurity system is obtained and the behavior of the correlation function exponents near its vicinity is elucidated.

\section{System description}

The system under study consists of a Luttinger liquid with short ranged mutual interactions amongst the particles and a cluster of impurities centered around an origin. The Hamiltonian of the system is given as follows.
\small
\begin{equation}
\begin{aligned}
H =& \int^{\infty}_{-\infty} dx \mbox{    } \psi^{\dagger}(x) \left( - \frac{1}{2m} \partial_x^2 + V(x) \right) \psi(x)\\
  & \hspace{1cm} + \frac{1}{2} \int^{ \infty}_{-\infty} dx \int^{\infty}_{-\infty} dx^{'} \mbox{  }v(x-x^{'}) \mbox{   }
 \rho(x) \rho(x^{'})
\label{Hamiltonian}
\end{aligned}
\end{equation}
\normalsize
The first term is the kinetic term followed by the potential energy term which represents the impurity cluster which is modeled as a finite sequence of barriers and wells around a fixed point. The potential cluster can be as simple as one delta impurity $V_0\delta(x)$ or two delta impurities placed close to each other $V_0( \delta(x+a)+\delta(x-a))$, finite barrier/well $\pm V \theta(x+a)\theta(a-x)$ and so on, where $\theta(x)$ is the Heaviside step function. The RPA (random phase approximation) is imposed on the system, without which the calculation of the analytical expressions of the correlation functions is formidable. In this limit, the Fermi momentum and the mass of the fermion are allowed diverge in such a way that their ratio, viz., the Fermi velocity is finite (i.e. $ k_F, m \rightarrow \infty $ but $ k_F/m = v_F < \infty  $). Under the choice of units: $ \hbar = 1 $, $ k_F $ is both the Fermi momentum as well as a wavenumber \cite{stone1994bosonization}. The RPA limit linearizes the energy momentum dispersion near the Fermi surface ($E=E_F+p v_F$ instead of $E=p^2/(2m)$). It is also imperative to define how the width of the impurity cluster `2a' scales in the RPA limit and the assertion is that  $ 2 a k_F   < \infty $ as $ k_F \rightarrow \infty $.  On the other hand, the heights and depths of the various barriers/wells are assumed to be in fixed ratios with the Fermi energy $ E_F = \frac{1}{2} m v_F^2 $ even as $ m \rightarrow \infty $ with $ v_F < \infty $. 

In case of the different potentials consisting the cluster, the only quantities that will be used in the calculation of the Green functions is the reflection (R) and transmission (T) amplitudes which can be easily calculated using elementary quantum mechanics and are provided in an earlier work \cite{das2018quantum}. For instance, in the case of a single delta potential: $V_0\delta(x)$,
\scriptsize
\begin{equation}
\begin{aligned}
T=&\frac{1}{\left(1+V_0 \frac{i}{v_F}\right)}\mbox{ };\mbox{ }
R=-\frac{iV_0}{v_F\left(1+V_0 \frac{i}{v_F}\right)} \\
\end{aligned}
\end{equation}
\normalsize
In the case of a double delta potential separated by a distance 2a between them : $V_0( \delta(x+a)+\delta(x-a))$,
\scriptsize
\begin{equation}
\begin{aligned}
T=&\frac{1}{\left(1+V_0 \frac{i}{v_F}\right)^2-\left(\frac{i V_0}{v_F}e^{i 2 k_F a}\right)^2}\\
R=&-\frac{2i\frac{V_0^2}{v_F^2} \sin{[2 k_F a]} +\frac{2i V_0}{v_F}\cos{[2 k_F a]}}{\left(1+V_0 \frac{i}{v_F}\right)^2-\left(\frac{i V_0}{v_F}e^{i 2 k_F a}\right)^2} \\
\end{aligned}
\end{equation}
\normalsize
In this work the generalized notion of R and T is used in this work to signify the reflection and transmission amplitudes of the cluster of impurities in consideration. The third term in equation (\ref{Hamiltonian}) represents the forward scattering mutual interaction term such that
\[ 
\hspace{2 cm} v(x-x^{'}) = \frac{1}{L} \sum_{q}  v_q \mbox{ }e^{ -i q(x-x^{'}) } 
\]
where $ v_q = 0 $ if $ |q| > \Lambda $ for some fixed bandwidth $ \Lambda \ll k_F $ and $ v_q = v_0 $ is a constant, otherwise.\\

\section{ Non chiral bosonization and two point functions}
As in conventional bosonization schemes using the operator approach \cite{giamarchi2004quantum}, the fermionic field operator is expressed in terms of currents and densities. But in NCBT the field operator is modified to include the effect of back-scattering by impurities. Hence it is suitable to study translationally non invariant systems like the ones considered in this work.
\begin{equation}
\begin{aligned}
\psi_{\nu}(x,\sigma,t) \sim C_{\lambda  ,\nu,\gamma}\mbox{ }e^{ i \theta_{\nu}(x,\sigma,t) + 2 \pi i \lambda \nu  \int^{x}_{sgn(x)\infty}\mbox{ } \rho_s(-y,\sigma,t) dy}
\label{PSINU}
\end{aligned}
\end{equation}
Here $\theta_{\nu}$ is the local phase which is a function of the currents and densities which is also present in the conventional bosonization schemes \cite{giamarchi2004quantum}, ideally suited for homogeneous systems.
\small
\begin{equation}
\begin{aligned}
\theta_{\nu}(x,\sigma,t) =& \pi \int^{x}_{sgn(x)\infty} dy \bigg( \nu  \mbox{  } \rho_s(y,\sigma,t)\\
&\hspace{1 cm} -  \int^{y}_{sgn(y)\infty} dy^{'} \mbox{ }\partial_{v_F t }  \mbox{ }\rho_s(y^{'},\sigma,t) \bigg)
\end{aligned}
\end{equation}\normalsize
The new addition in equation (\ref{PSINU}) is the optional term $\rho_s(-y)$ which ensures the necessary trivial exponents for the single particle Green functions for a system of otherwise free fermions with impurities, which are obtained using standard Fermi algebra and they serve as a basis for comparison for the Green functions obtained using bosonization. The adjustable parameter is the quantity $\lambda$ which can take values either 0 or 1 as per requirement. Thus NCBT operator reduces to standard bosonization operator used in g-ology methods by setting $\lambda=0$. The factor $2 \pi i$ ensures that the fermion commutation rules are obeyed. The quantities $C_{\lambda  ,\nu,\gamma}$ are pre-factors and are fixed by comparison with the non-interacting Green functions obtained using Fermi algebra. The suffix $\nu$ signifies a right mover or a left mover and takes values 1 and -1 respectively. The field operator as given in equation (\ref{PSINU}) is to be treated as a mnemonic to obtain the Green functions and not as an operator identity, which avoids the necessity of the Klein factors that are conventionally used to conserve the number as the correlation functions, unlike the field operators, are number conserving. The field operator (annihilation) is clubbed together with another such field operator (creation) to obtain the non interacting two point functions after fixing the C's and $\lambda$'s. Finally the densities $\rho$'s in the RHS of equation (\ref{PSINU}) are replaced by their interacting versions to obtain the many body Green functions, the details being described in an earlier work \cite{das2018quantum}. The two point functions obtained using NCBT are given in \hyperref[AppendixA]{Appendix A}.

\section{Conductance }
\subsection{Kubo conductance}
The general formula for the conductance of a quantum wire (obtained from Kubo's formula that relates it to current-current correlations) without leads but with electrons experiencing forward scattering short-range mutual interactions
and in the presence of a finite number of barriers and wells clustered around an origin is obtained.
Consider an electric field $ E(x,t)  = \frac{ V_g }{ L}  $ between $ -\frac{L}{2} < x < \frac{L}{2} $
and $ E(x,t) = 0 $ for $ |x| > \frac{L}{2} $. Here $ V_g $ is the voltage between two extreme points.
Thus a d.c. situation is being considered right from the start. This corresponds to a vector potential,
\begin{equation}
\begin{aligned}
A(x,t) = \left\{
  \begin{array}{ll}
   -\frac{ V_g }{ L} (ct), & \hbox{ $  -\frac{L}{2} < x < \frac{L}{2} $ ;} \\
   \hspace{.3cm}0, & \hbox{otherwise.}
  \end{array}
\right.
\end{aligned}
\end{equation}
Here c is the speed of light. This means the average current can be written as,
\begin{equation}
\begin{aligned}
<j(x,\sigma,t)>  =  &\frac{ie}{c}\sum_{ \sigma^{'} }
\int^{L/2}_{-L/2} dx^{'}\mbox{  }  \int_{-\infty}^{t} dt^{'}
\mbox{  }\frac{ V_g }{ L} (ct') \\
&< [j(x,\sigma,t),j(x^{'},\sigma^{'},t^{'})]>_{LL}
\end{aligned}
\end{equation}
The current current correlation can be obtained using the Green functions derived in the present work (see \hyperref[AppendixB]{Appendix B}) to obtain the formula for conductance (in proper units) as follows,
\begin{equation}
G = \frac{ e^2 }{h}    \frac{v_F  }{ v_h  } \mbox{   }\bigg (1- \frac{v_F }{v_h}  \mbox{    }\frac{|R|^2}{1-\frac{(v_h-v_F)}{v_h}|R|^2}\bigg)
\label{kubo}
\end{equation}

Here $v_F$ is the Fermi velocity, \scriptsize $ v_h = \sqrt{v_F^2+2v_F v_0/\pi} $ \normalsize is the holon velocity and $v_0$ is the strength of interaction between fermions as already described in Section 2. See \hyperref[AppendixB]{Appendix B} for more details.
\begin{figure}[h!]
  \centering
  \includegraphics[scale=0.3]{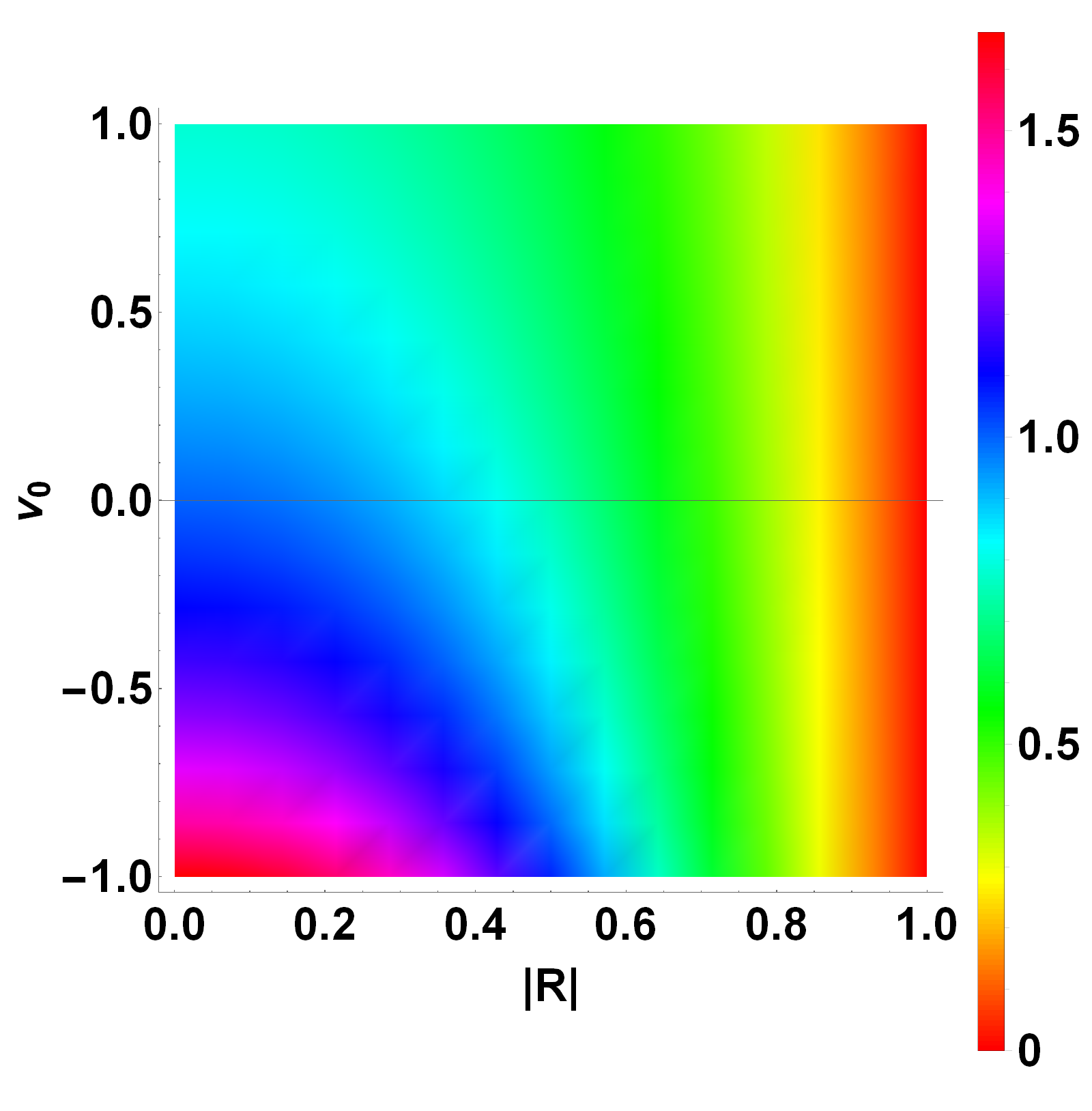}
  \caption{Conductance as a function of the absolute value of the reflection amplitude as well the interaction parameter ($ v_F = 1 $)}\label{Cond3D}
\end{figure}
\noindent The Kubo conductance formula obtained in equation (\ref{kubo}) is plotted in fig. \ref{Cond3D} as a function of the reflection coefficient and interaction strength.  It can be seen that when the reflection coefficient becomes unity ($|R|=1$), then the conductance vanishes irrespective of the interaction parameter. On the other hand, for any fixed value of $|R|$, the conductance increases as the mutual  interaction becomes more and more attractive (negative $v_0$) and decreases as the interaction becomes more and more repulsive (positive $v_0$). On the other hand for a fixed value of interaction parameter, the conductance decreases with increase in the reflection parameter.

\subsubsection{Limiting cases.}
{\bf No interaction}. In absence of interactions $v_0=0$ and hence $v_h=v_F$ and thus from equation (\ref{kubo}),
\[
\hspace{2cm}G = \frac{ e^2 }{h} (1 - |R|^2) = \frac{ e^2 }{h} |T|^2
\]
 which is the Landauer's formula for conductance.

{\bf No impurity} In this case, there is no reflection and hence $|R|=0$ and thus from equation (\ref{kubo}),
\[
\hspace{2cm}G = \frac{ e^2 }{h} \frac{v_F}{v_h} =\frac{ e^2 }{h} g
\]
which the renormalized conductance of an infinite Luttinger liquid (with parameter g).

{\bf Infinite barrier}
In the case of a half line, $|R|=1$ and thus from equation (\ref{kubo}),
\[
\hspace{1 in}G=0
\]
irrespective of the value of holon velocity $v_h$.\\

\subsection{Tunneling conductance} The Kubo conductance is the linear response to external potentials and is therefore related to four-point correlation functions of fermions. Alternatively, conductance may also be thought of the outcome of a tunneling experiment \cite{kane1992transport}.
Here fermions are injected from one end and collected from the other end. In this sense the conductance is related to the two-point function or the single particle Green function. Thus we expect these two notions to be qualitatively different from each other.  From this point of view, the conductance is ($|T| $ is the magnitude of the transmission amplitude for free fermions plus impurity) ,
\begin{equation}
G = \frac{ e^2 }{h } |T| \mbox{   }
| v_F\int^{\infty}_{-\infty}dt\mbox{   }<\{ \psi_{ R } ( \frac{L}{2},\sigma,t) ,  \psi^{\dagger}_{ R } (-\frac{L}{2},\sigma,0)  \}>
 |
 \label{TUNNEL1}
\end{equation}
In this case the results depend on the length of the wire $ L $ and a cutoff $ L_{\omega} = \frac{ v_F }{ k_B T } $ that may be regarded either as inverse temperature or inverse frequency (in case of a.c. conductance). The result (derived in \hyperref[AppendixB]{Appendix B}) is
\begin{equation}
G \sim \left( \frac{ L }{ L_{ \omega } }\right)^{-2Q }  \mbox{  }  \left( \frac{ L }{ L_{ \omega} }\right)^{ 4X }
\label{GGEN}
\end{equation}
Here Q and X are obtained from equation (\ref{luttingerexponents}). It is important to stress that the present work has carefully defined tunneling conductance and it is not simply related to the dynamical density of states of either the bulk or the half line (\hyperref[AppendixB]{Appendix B}). Of particular interest is the weak link limit where $ |R| \rightarrow 1 $. The limiting case of the weak link are two semi-infinite wires.
In this case,
\begin{equation}
G_{weak-link} \sim \left( \frac{ L }{ L_{ \omega} }\right)^{ \frac{ (v_h + v_F)^2-4v^2_F }{ 4 v_h v_F } }
\label{GGEN}
\end{equation}
Hence the d.c. conductance scales as $ G_{weak-link} \sim (k_B T)^{ \frac{ (v_h + v_F)^2-4v^2_F }{ 4 v_h v_F } } $.  This formula is consistent with the
assertions of Kane and Fisher \cite{kane1992transport} that show that at low temperatures $ k_B T \rightarrow 0 $
for a fixed $ L $, the conductance vanishes as a power law in the temperature if the interaction between the fermions is repulsive ($ v_h > v_F  > 0 $) and diverges as a power law if the interactions between the fermions is attractive ($ v_F > v_h > 0 $). Their result is applicable to spinless fermions without leads $ G_{weak-link-nospin} \sim (k_B T)^{ \frac{2}{K} - 2 } $. In order to compare with the result of the present work, this exponent has to be halved $  G_{weak-link-with-spin} \sim (k_B T)^{ \frac{1}{K_{ \rho } } - 1 }  $. This exponent is the same as the exponent of the present work so long as $ |v_h-v_F| \ll v_F $ ie. $ \frac{ (v_h + v_F)^2-4v^2_F }{ 4 v_h v_F } \approx \frac{1}{K_{ \rho } } - 1 $ since $ K_{\rho} = \frac{ v_F }{v_h} $.
%
\begin{figure}[h!]
  \centering
 \includegraphics[scale=0.35]{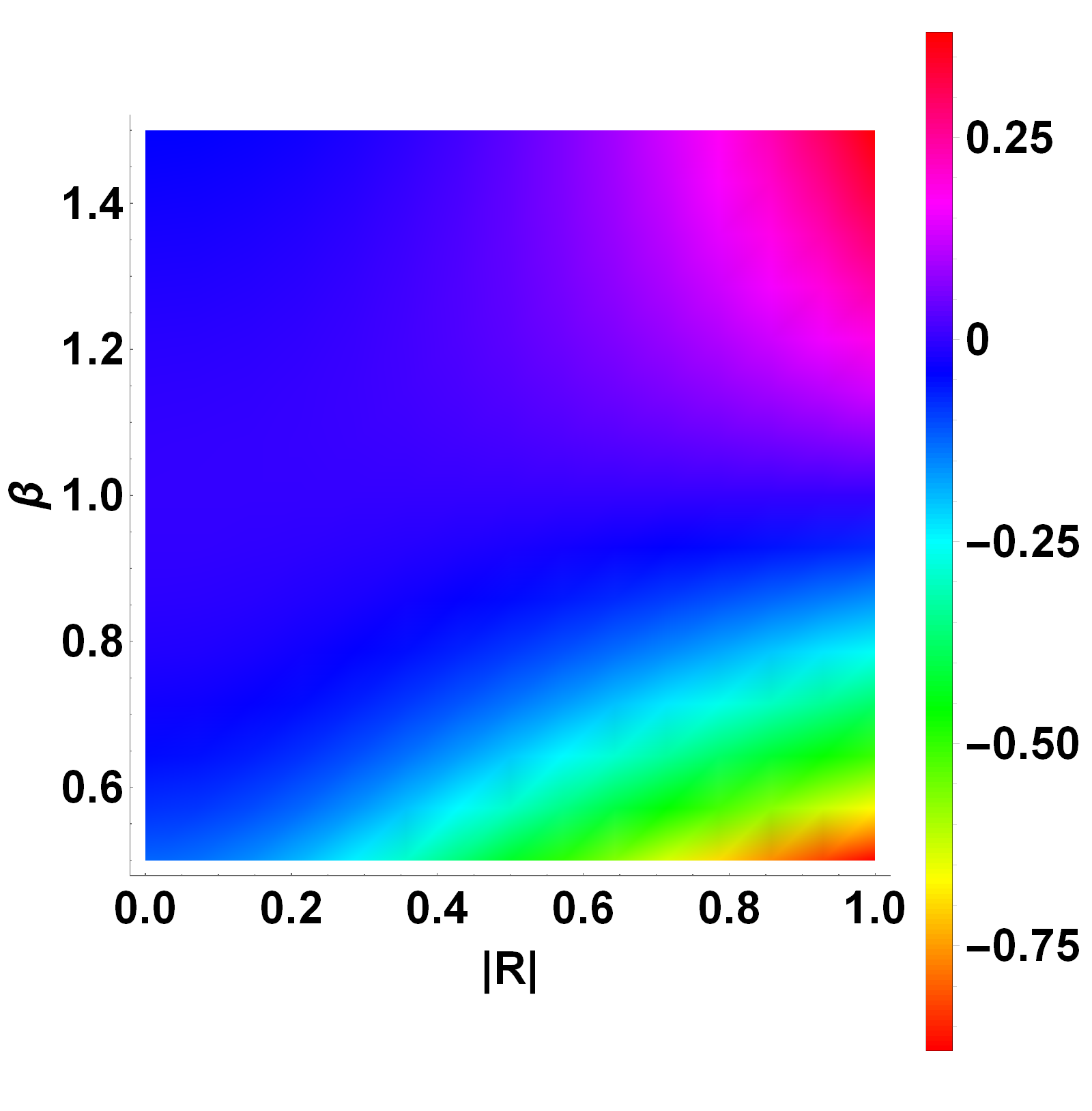}
  \caption{Conductance exponent $\eta$ as a function of the absolute value of the reflection amplitude $|R|$ and the ratio $\beta=\frac{v_h}{v_F}$. }\label{eta}
\end{figure}
%
In general, the claim of the present work  is that the temperature dependence of the tunneling d.c. conductance of a wire with no leads and in the presence of barriers and wells and mutual interaction between particles (forward scattering, infinite bandwidth ie. $ k_F \gg \Lambda_b \rightarrow \infty $) is,
\begin{equation}
G \sim (k_B T)^{ \eta} ;\mbox{   }\mbox{  } \mbox{  } \eta = 4X - 2 Q
\label{Cond}
\end{equation}
When $ \eta > 0 $ the conductance vanishes at low temperatures as a power law - characteristic of a weak link. However when $ \eta < 0 $ the conductance diverges at low temperature as a power law - characteristic of a clean quantum wire. Of special interest is the situation $ \eta = 0 $ where the  conductance is independent of temperature. This crossover from a conductance that vanishes as a power law at low temperatures to one that diverges as a power law occurs at reflection coefficient
 $ |R|^2 = |R_{c2}|^2 \equiv \frac{v_h (v_h-v_F)}{3 v_F^2+v_h^2} $ which is valid only for repulsive interactions $ v_h > v_F $. For attractive interactions, $ \eta < 0 $ for any $ |R|^2 $ which means
 the conductance always diverges as a power law at low temperatures. This means attractive interactions heal the chain for all reflection coefficients including in the extreme weak link case.
 On the other hand for repulsive interactions, for $ |R| > |R_{c2}| $, $ \eta > 0 $ the chain is broken (conductance vanishes) at low temperatures. For $ |R| < |R_{c2}| $, $ \eta < 0 $ and even though the interactions are repulsive the chain is healed (conductance diverges).

\begin{figure*}[t!]
\begin{center}
\includegraphics[scale=0.12]{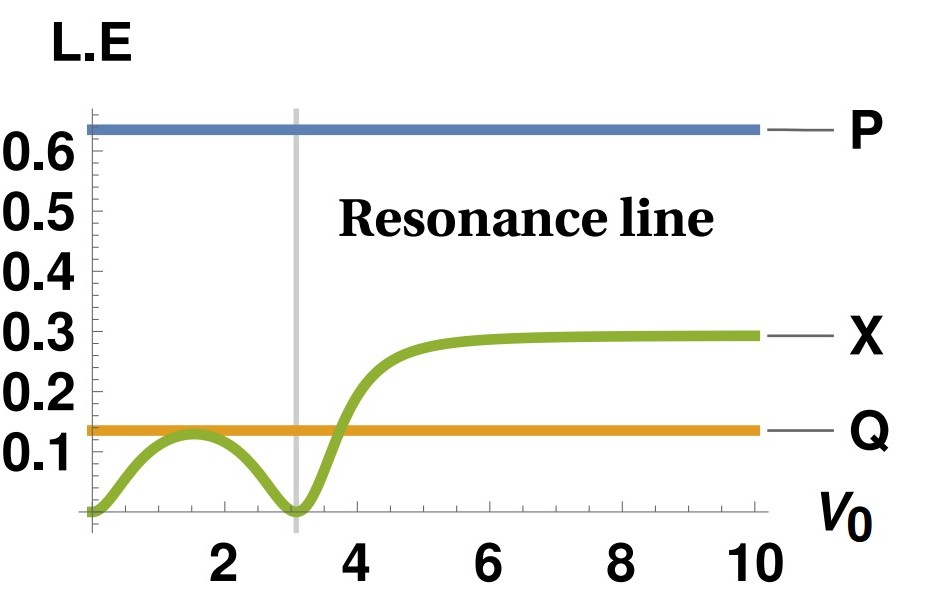}\hspace{0.5 cm}
\includegraphics[scale=0.17]{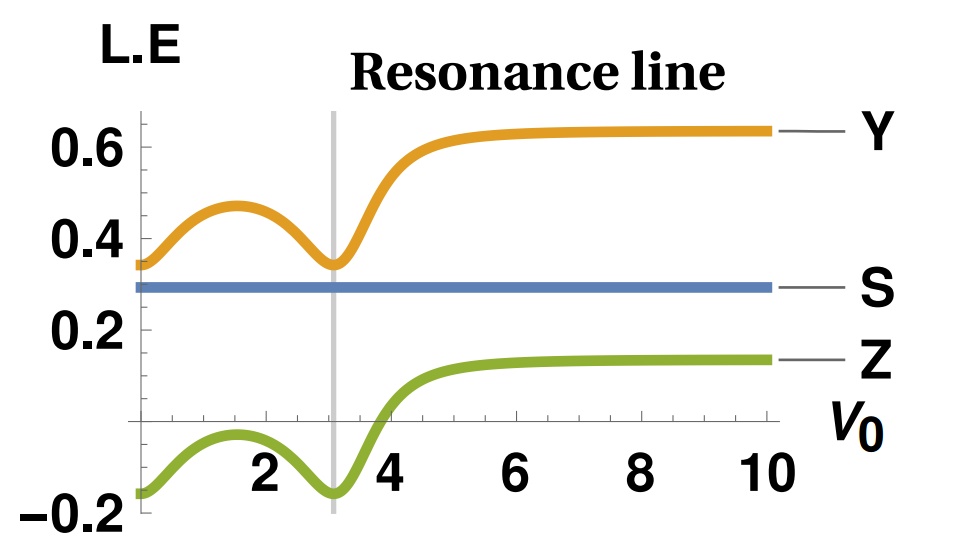}\hspace{0.5 cm}
\includegraphics[scale=0.17]{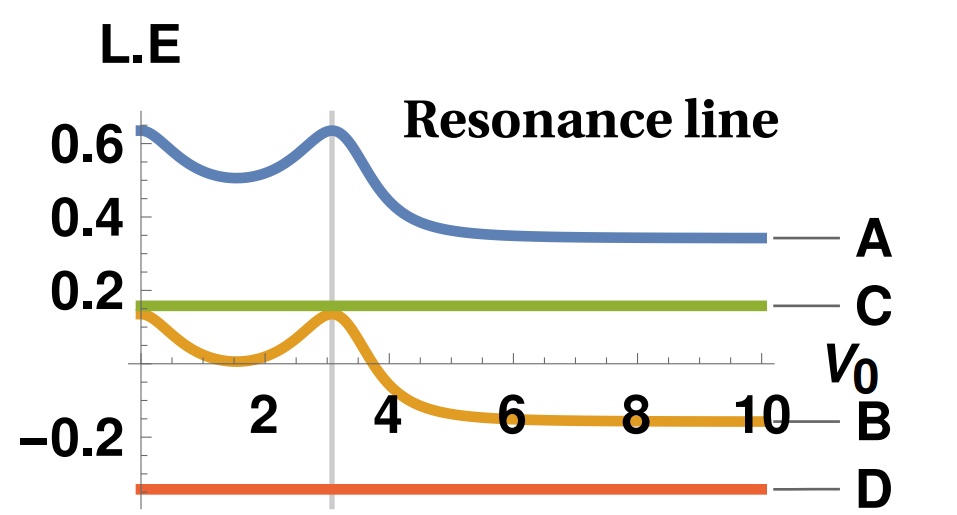}

\scriptsize (a) \hspace{5 cm}(b)\hspace{5 cm} (c)
\end{center}
\caption{ Anomalous exponents (L.E) vs impurity strength $V_0$ for symmetric double barrier: (a) Exponents for $\langle \psi_R(X_1) \psi_R^{\dagger}(X_2)\rangle$ on the same side (b) Exponents for $\langle \psi_R(X_1) \psi_L^{\dagger}(X_2)\rangle$ on the same side  (c) Exponents for $\langle \psi_R(X_1) \psi_R^{\dagger}(X_2)\rangle$ on opposite sides.}
\label{resonance}
\end{figure*}

\subsubsection{ Derivation of RG equation for the tunneling conductance }

In the well-cited work of Matveev et al \cite{matveev1993tunneling}, the RG equation for the tunneling conductance is derived which is valid for weak mutual interaction between fermions (they consider both forward scattering as well as backward scattering but in the present work we consider only forward scattering between fermions but of arbitrary strength and sign subject to the limitation that the holon velocity be real). Both in their work and in the present work the transmission amplitude of free fermions can vary continuously between zero and unity i.e. it is not constrained in any way. Note that we have chosen an infinite bandwidth to derive the power-law conductance in equation (\ref{Cond}). Had we chosen a finite bandwidth while calculating equation (\ref{TUNNEL1}), the resulting expressions would be considerably more complicated as Matveev  et al have also found. We shall postpone a proper discussion of this interesting question to a later publication. For now we look at equation (8) of their paper rather than equation (12) since we are interested in the large bandwidth case only for now. Since $ G \sim {\mathcal{T}} $ in their notation, we may expand the conductance exponent $ 4X - 2 Q $ in powers of $ v_0 $ the forward scattering mutual interaction between fermions to leading order (in the notation of Matveev et al this is $ V(0) $ and $ V(2k_F) \equiv 0 $ in the present work),
\begin{equation}
\frac{ \delta {\mathcal{T}} }{  {\mathcal{T}}_0 } \approx 4 X \mbox{ } \log(\omega)\approx {\mathcal{R}}_0\frac{ v_0 }{ \pi v_F} \mbox{ } \log(\omega)
\label{Matveev}
\end{equation}
for $ |v_0| \ll v_F $.
where $ {\mathcal{R}}_0 = 1 - {\mathcal{T}}_0 $ (in the notation of the present work this would be $ |R|^2 = 1- |T|^2 $ and $  \omega  \rightarrow |k-k_F|d \sim k_BT $. The equation (\ref{Matveev}) is precisely equation (8) of Matveev et al. Thus mutually interacting fermions renormalize the impurities but isolated impurities do not renormalize the homogenous Luttinger parameters such as $ K = \frac{v_F}{v_h} $. Note that our results for the conductance equation (\ref{Cond}) is the {\it{ end result }} of properly taking into account the renormalizations to all orders in the infinite-bandwidth-forward-scattering fermion-fermion interactions with no restriction on the bare transmission coefficient of free fermions plus impurity. The final answers of equation (\ref{Cond}) involve only the bare transmission and reflection coefficients for the same reason why the zero point energy of the harmonic oscillator derived properly using Hermite polynomials (rather than using perturbative RG around free particle, say) involves the bare spring constant (ie. $ \frac{1}{2} \hbar \sqrt{\frac{ k }{m } } $). Incidentally, even the final answers of Matveev et al. such as their equation (13) involve the bare parameters only since this formula is the {\it{end result}} of taking into account all the renormalization properly.

 It is hard to overstate the importance of these results. They show that it is possible to analytically interpolate between the weak barrier and weak link limits without involving RG techniques. It also shows that NCBT is nothing but non-perturbative RG in disguise.

\section{Resonant tunneling across a double barrier}
\begin{figure*}[t!]
\begin{center}
\includegraphics[scale=0.45]{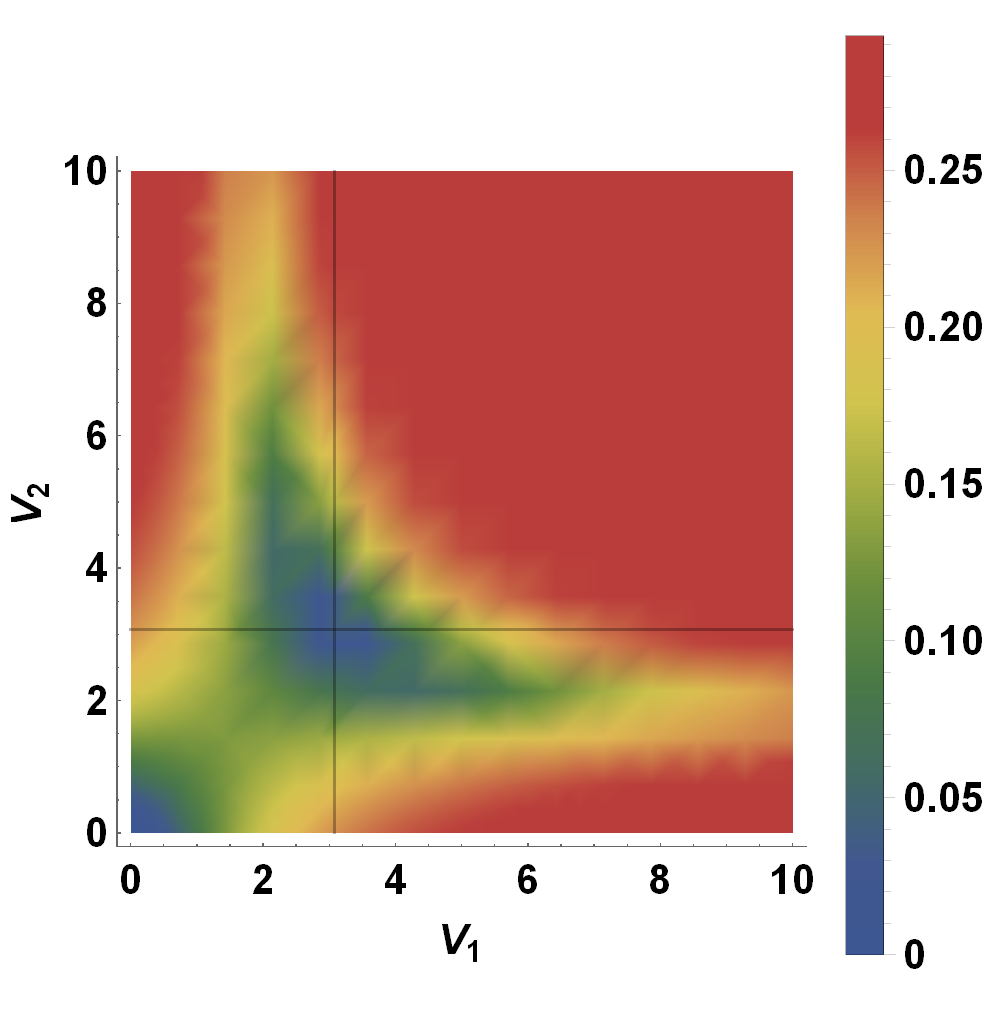}\hspace{1.5cm}
\includegraphics[scale=0.45]{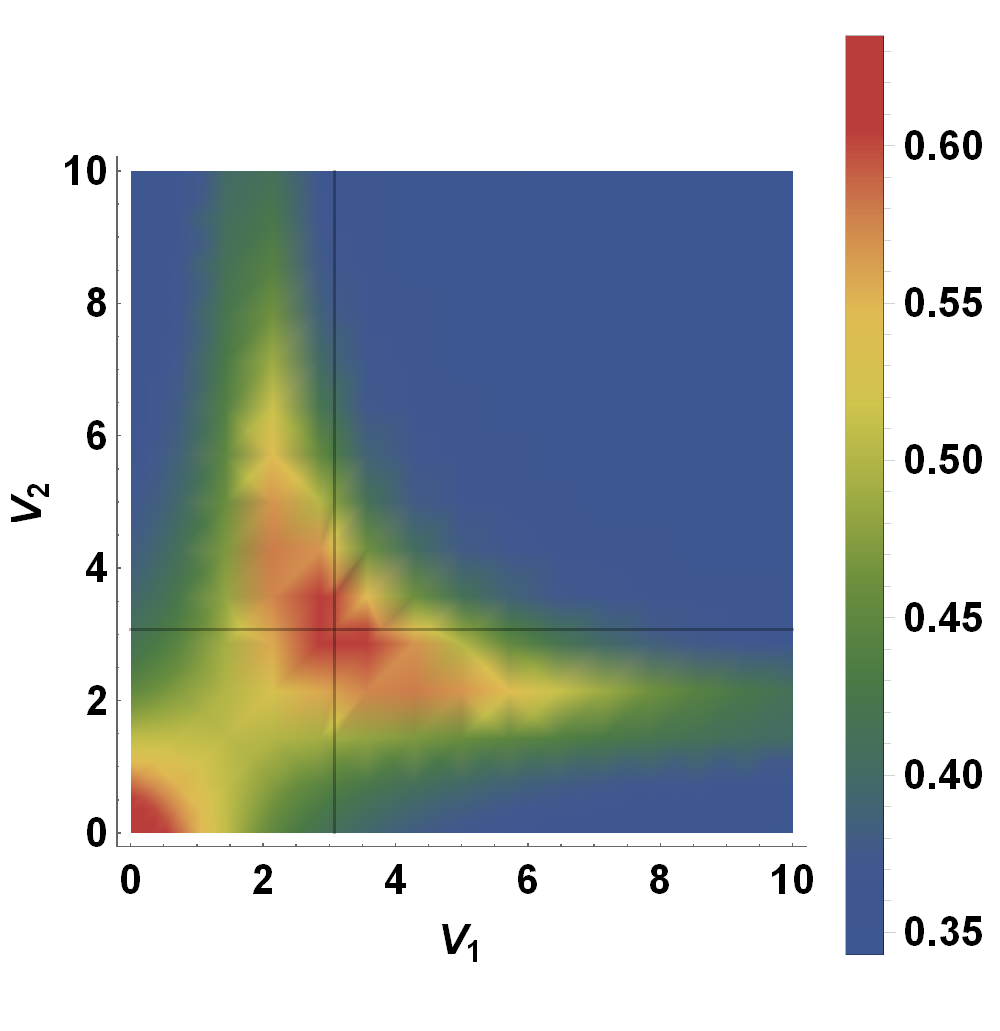}\\
(a)\hspace{7cm}(b)\\
\end{center}
\caption{ Anomalous exponents for double barrier: The anomalous exponents (a) X and (b) A  as functions of impurity strength $V_1$ and $V_2$ for an asymmetric double  delta  potential. Near  resonance (the point of intersection of the cross lines), the system has the same colour it has when both $V_1$ and $V_2$ are zero.}
\label{densityplot}
\end{figure*}

Resonant tunneling is well-known in elementary quantum mechanics. Typically, this phenomenon is studied in a double-barrier system. When the Fermi wavenumber bears a special relation with the inter-barrier separation and height, the reflection coefficient becomes zero and the Green functions of the system behave as if they are those of a translationally invariant system.  Consider a symmetric double delta-function with strength $V_0 $ and separation $ d $. Define, $ \xi_0 = k_F d $. The resonance condition in this case is well-known to be,

\begin{equation}
\hspace{0.5 in }V_0 \sin{[\xi_0]} +v_F\cos{[\xi_0]}=0  \label{eq:cond}
\end{equation}
Resonant tunneling is studied for a square double barrier potential in one dimensions by Zhi Xiao et al. \cite{xiao2012revisiting}. After taking the limiting cases of the square barriers tending to delta potentials and imposing the RPA limit, equation (\ref{eq:cond}) is obtained.

The anomalous exponents of the correlation functions given in \hyperref[AppendixA]{Appendix A} are plotted in fig. \ref{resonance} in the vicinity of resonance to see the signatures of resonance tunneling on the Luttinger liquid Green function.
It may be  seen that when the system is at resonance (depicted by the vertical line), all the anomalous exponents take exactly the same value that they take when there is no barrier at all.\\

For an asymmetric double delta system, $V(x)=V_1 \delta(x+a)+ V_2 \delta(x-a)$, the anomalous exponents can be calculated using NCBT. The form of the exponents are the same as given in \hyperref[AppendixA]{Appendix A} but the expression of the reflection amplitude is now different and is given by (here $\xi_0= 2 k_F a$) \cite{das2018quantum}.
\begin{equation}
\begin{aligned}
\label{asymmetric}
R=&-\frac{2 i \frac{V_1 V_2}{v_F^2} \sin{[\xi_0]}+\frac{2i}{v_F}(\frac{V_1 e^{i \xi_0}+V_2e^{-i\xi_0}}{2})}{\left(1+i\frac{V_1+V_2}{v_F}+\frac{i^2 V_1V_2}{v_F^2}\right)+\frac{V_1V_2}{v_F^2}e^{2 i \xi_0}}\\
\end{aligned}
\end{equation}
 For this case also resonance is achieved when both $V_1$ and $V_2$ becomes equal    ($V_1=V_2=V_0$) and $V_0$ obeys the same condition in equation $(\ref{eq:cond})$. Two of the anomalous exponents X and A (expressions given in equations (\ref{luttingerexponents}) and (\ref{asymmetric})) for the asymmetric double delta system are plotted in fig. (\ref{densityplot}). The point of intersection of the cross lines is the condition for resonance and it can easily be seen that the exponent takes the same value (color) at resonance point as it otherwise takes for the no-impurity system ($V_1=V_2=0$). 

\section{Conclusion}
The correlation functions of an inhomogeneous Luttinger liquid obtained using the Non chiral bosonization are successfully used to calculate the conductance in the Kubo formalism as well as in a tunneling experiment. The formulas are valid for any strength of the impurities as well as that of the inter-particle interactions and various standard results are obtained as limiting cases of these formulas. The condition of resonant tunneling is also obtained and the behavior of the correlation functions near resonance is described.

\section*{APPENDIX A:  Two point functions using NCBT}
\label{AppendixA}
\setcounter{equation}{0}
\renewcommand{\theequation}{A.\arabic{equation}}

The full Green function is the sum of all the parts. The notion of weak equality is introduced which is denoted by \begin{small} $ A[X_1,X_2] \sim B[X_1,X_2] $ \end{small}. This really means  \begin{small} $ \partial_{t_1} Log[ A[X_1,X_2] ]  = \partial_{t_1} Log[ B[X_1,X_2] ] $\end{small} assuming that A and B do not vanish identically. In addition to this, the finite temperature versions of the formulas below can be obtained by replacing $ Log[Z] $ by $ Log[ \frac{\beta v_F }{\pi}Sinh[ \frac{\pi Z}{ \beta v_F} ] ] $ where $ Z \sim  (\nu x_1 - \nu^{'} x_2 ) - v_a (t_1-t_2)  $ and singular cutoffs ubiquitous in this subject are suppressed in this notation for brevity - they have to be understood to be present. {\bf Notation:} $X_i \equiv (x_i,\sigma_i,t_i)$ and  $\tau_{12} =  t_1 - t_2$. 
\scriptsize

\begin{equation}
\begin{aligned}
\Big\langle T\mbox{  }\psi(X_1)\psi^{\dagger}(X_2) \Big\rangle 
=&\Big\langle T\mbox{  }\psi_{R}(X_1)\psi_{R}^{\dagger}(X_2) \Big\rangle +\Big \langle T\mbox{  }\psi_{L}(X_1)\psi_{L}^{\dagger}(X_2)\Big\rangle \\
+&\Big\langle T\mbox{  }\psi_{R}(X_1)\psi_{L}^{\dagger}(X_2) \Big\rangle + \Big\langle T\mbox{  }\psi_{L}(X_1)\psi_{R}^{\dagger}(X_2)\Big\rangle \\
\label{break}
\end{aligned}
\end{equation}

\small
\begin{bf} Case I : $x_1$ and $x_2$ on the same side of the origin\end{bf} \\ \scriptsize

\begin{equation*}
\begin{aligned}
\Big\langle T\mbox{  }\psi&_{R}(X_1)\psi_{R}^{\dagger}(X_2)\Big\rangle \sim 
\frac{(4x_1x_2)^{\gamma_1}}{(x_1-x_2 -v_h \tau_{12})^{P} (-x_1+x_2 -v_h \tau_{12})^{Q}} \\
\times&\frac{1}{ (x_1+x_2 -v_h \tau_{12})^{X} (-x_1-x_2 -v_h \tau_{12})^{X} (x_1-x_2 -v_F \tau_{12})^{0.5}}\\
\Big\langle T\mbox{  }\psi&_{L}(X_1)\psi_{L}^{\dagger}(X_2)\Big\rangle \sim 
\frac{(4x_1x_2)^{\gamma_1}}{(x_1-x_2 -v_h \tau_{12})^{Q} (-x_1+x_2 -v_h \tau_{12})^{P}} \\
\times&\frac{1}{ (x_1+x_2 -v_h \tau_{12})^{X} (-x_1-x_2 -v_h \tau_{12})^{X}(-x_1+x_2 -v_F \tau_{12})^{0.5}}\\
\Big\langle T\mbox{  }\psi&_{R}(X_1)\psi_{L}^{\dagger}(X_2)\Big\rangle \sim 
\frac{(2x_1)^{\gamma_1}(2x_2)^{1+\gamma_2}+(2x_1)^{1+\gamma_2}(2x_2)^{\gamma_1}}{2(x_1-x_2 -v_h \tau_{12})^{S} (-x_1+x_2 -v_h \tau_{12})^{S}} \\
\times&\frac{1}{ (x_1+x_2 -v_h \tau_{12})^{Y} (-x_1-x_2 -v_h \tau_{12})^{Z}(x_1+x_2 -v_F \tau_{12})^{0.5}}\\
\end{aligned}
\end{equation*}

\begin{equation}
\begin{aligned}
\Big\langle T\mbox{  }\psi&_{L}(X_1)\psi_{R}^{\dagger}(X_2)\Big\rangle \sim 
\frac{(2x_1)^{\gamma_1}(2x_2)^{1+\gamma_2}+(2x_1)^{1+\gamma_2}(2x_2)^{\gamma_1}}{2(x_1-x_2 -v_h \tau_{12})^{S} (-x_1+x_2 -v_h \tau_{12})^{S}} \\
\times&\frac{1}{ (x_1+x_2 -v_h \tau_{12})^{Z} (-x_1-x_2 -v_h \tau_{12})^{Y}(-x_1-x_2 -v_F \tau_{12})^{0.5}}\\
\label{SS}
\end{aligned}
\end{equation}

\small
\begin{bf}Case II : $x_1$ and $x_2$ on opposite sides of the origin\end{bf} \\ \scriptsize

\begin{equation}
\begin{aligned}
\Big\langle T\mbox{  }\psi&_{R}(X_1)\psi_{R}^{\dagger}(X_2)\Big\rangle \sim 
\frac{(2x_1)^{1+\gamma_2}(2x_2)^{\gamma_1} }{2(x_1-x_2 -v_h \tau_{12})^{A} (-x_1+x_2 -v_h \tau_{12})^{B}} \\
\times&\frac{(x_1+x_2)^{-1}(x_1+x_2 + v_F \tau_{12})^{0.5}}{ (x_1+x_2 -v_h \tau_{12})^{C} (-x_1-x_2 -v_h \tau_{12})^{D} (x_1-x_2 -v_F \tau_{12})^{0.5}}\\
&\hspace{2cm}+\frac{(2x_1)^{\gamma_1} (2x_2)^{1+\gamma_2}}{2(x_1-x_2 -v_h \tau_{12})^{A} (-x_1+x_2 -v_h \tau_{12})^{B}} \\
\times&\frac{(x_1+x_2)^{-1}(x_1+x_2 - v_F \tau_{12})^{0.5}}{ (x_1+x_2 -v_h \tau_{12})^{D} (-x_1-x_2 -v_h \tau_{12})^{C} (x_1-x_2 -v_F \tau_{12})^{0.5}}\\
\Big\langle T\mbox{  }\psi&_{L}(X_1)\psi_{L}^{\dagger}(X_2)\Big\rangle \sim 
\frac{(2x_1)^{1+\gamma_2}(2x_2)^{\gamma_1} }{2(x_1-x_2 -v_h \tau_{12})^{B} (-x_1+x_2 -v_h \tau_{12})^{A}} \\
\times&\frac{(x_1+x_2)^{-1}(x_1+x_2 - v_F \tau_{12})^{0.5}}{ (x_1+x_2 -v_h \tau_{12})^{D} (-x_1-x_2 -v_h \tau_{12})^{C} (-x_1+x_2 -v_F \tau_{12})^{0.5}}\\
&\hspace{2cm}+\frac{(2x_1)^{\gamma_1} (2x_2)^{1+\gamma_2}}{2(x_1-x_2 -v_h \tau_{12})^{B} (-x_1+x_2 -v_h \tau_{12})^{A}} \\
\times&\frac{(x_1+x_2)^{-1}(x_1+x_2 + v_F \tau_{12})^{0.5}}{ (x_1+x_2 -v_h \tau_{12})^{C} (-x_1-x_2 -v_h \tau_{12})^{D} (-x_1+x_2 -v_F \tau_{12})^{0.5}}\\
\Big\langle T\mbox{  }\psi&_{R}(X_1)\psi_{L}^{\dagger}(X_2)\Big\rangle \sim \mbox{ }0\\
\Big\langle T\mbox{  }\psi&_{L}(X_1)\psi_{R}^{\dagger}(X_2)\Big\rangle \sim  \mbox{ }0\\
\label{OS}
\end{aligned}
\end{equation}
\normalsize
where
\footnotesize
\begin{equation}
Q=\frac{(v_h-v_F)^2}{8 v_h v_F} \mbox{ };\mbox{ }  X=\frac{|R|^2(v_h-v_F)(v_h+v_F)}{8  v_h (v_h-|R|^2 (v_h-v_F))}  \mbox{ };\mbox{ }C=\frac{v_h-v_F}{4v_h}
\label{luttingerexponents}\end{equation}
\normalsize
The other exponents can be expressed in terms of the above exponents.
\footnotesize
\begin{equation*}
\begin{aligned}
&P= \frac{1}{2}+Q  \mbox{ };\hspace{0.8 cm}    S=\frac{Q}{C}( \frac{1}{2}-C)   \mbox{ };\hspace{0.85 cm}      Y=\frac{1}{2}+X-C  ;           \\
& Z=X-C\mbox{ };\hspace{0.8 cm}      A=\frac{1}{2}+Q-X \mbox{ };\hspace{0.8 cm}   B=Q-X  \mbox{ };\hspace{1 cm}   \\
&D=-\frac{1}{2}+C   \mbox{ };\hspace{.6 cm}      \gamma_1=X                \mbox{ };\hspace{1.65 cm}    \gamma_2=-1+X+2C;\\
\end{aligned}
\end{equation*}
\normalsize
\section*{APPENDIX B:  Conductance of a quantum wire}
\label{AppendixB}
\setcounter{equation}{0}
\renewcommand{\theequation}{B.\arabic{equation}}

In this section, the conductance of a quantum wire with no leads is discussed first using Kubo's formula and next using the idea that it is the outcome of a tunneling experiment.
\subsection{Kubo formalism}
The electric field is $ E(x,t)  = \frac{ V_g }{ L}  $ between $ -\frac{L}{2} < x < \frac{L}{2} $ and $ E(x,t) = 0 $ for $ |x| > \frac{L}{2} $. Here $ V_g $ is the Voltage between two extreme points. Thus a d.c. situation is being considered right from the start. This corresponds to a vector potential ( c is the velocity of light),
\[
A(x,t) = \left\{
  \begin{array}{ll}
   -\frac{ V_g }{ L} (ct), & \hbox{ $  -\frac{L}{2} < x < \frac{L}{2} $ ;} \\
   0, & \hbox{otherwise.}
  \end{array}
\right.
\]
This means (since $ j \approx j_s $, the slow part) ,
\begin{equation}
\begin{aligned}
<j(x,\sigma,t)>  =  &\frac{ie}{c}\sum_{ \sigma^{'} }
\int^{L/2}_{-L/2} dx^{'}\mbox{  }  \int_{-\infty}^{t} dt^{'} \\
&\times\frac{ V_g }{ L} (ct') < [j(x,\sigma,t),j(x^{'},\sigma^{'},t^{'})]>_{LL}
\label{gencond}
\end{aligned}
\end{equation}

\subsubsection{ Clean wire: $  |R| = 0 $ but $ v_0 \neq 0 $ }
Using the Green function from equation (\ref{SS}) and setting $|R|=0$, the current current commutation relation can be calculated as,
\footnotesize
\begin{equation}
\begin{aligned}
<[j_s&(x,\sigma,t),j_s(x',\sigma',t')]> 
=-\frac{v^2_F  }{ 8\pi^2 } \mbox{   } \sum_{  \nu = \pm 1 }
 (2 \pi i) \\
&\partial_{ v_F t' }\left( \delta( x-x' + \nu v_h(t-t') ) + \sigma \sigma' \mbox{   }\delta( x-x' + \nu v_F(t-t') ) \right)
\label{cleancond}
\end{aligned}
\end{equation}
\normalsize
Inserting equation (\ref{cleancond}) into equation (\ref{gencond}), the following is obtained.\footnotesize
\begin{equation*}
\begin{aligned}
<j&(x,\sigma,t)>  =  \frac{ie}{c}\sum_{ \sigma^{'} }
\int^{\frac{L}{2}}_{-\frac{L}{2}} dx^{'} \int_{-\infty}^{t} dt^{'} \mbox{  }\frac{ V_g }{ L} (ct') 
\Big(\frac{-v^2_F  }{ 8\pi^2 } \mbox{   } \sum_{  \nu = \pm 1 }
 (2 \pi i) \mbox{  }\\
\times&\partial_{ v_F t' }\left( \delta( x-x' + \nu v_h(t-t') )  
\mbox{ }+ \sigma \sigma' 
\delta( x-x' + \nu v_F(t-t') ) \right)\Big)
\end{aligned}
\end{equation*}\normalsize
Finally,
\[
<j(x,\sigma,t)>  = -
\mbox{  }  V_g \frac{e }{ (2\pi) }  \frac{ v_F }{ v_h}
\]
or,
\[
I = (-e) <j(x,\sigma,t)>  =
\mbox{  }  V_g \frac{e^2 }{ (2\pi) }  \frac{ v_F }{ v_h}
\]
This gives the formula for the conductance (per spin) for a clean quantum wire with interactions,
\[
G = \frac{ e^2}{2\pi }  \frac{ v_F }{ v_h}
\]
\normalsize
or in proper units,
\[
\begin{boxed}
{G = \frac{ e^2}{2\pi \hbar}\mbox{  } \frac{ v_F }{ v_h}  = \frac{ e^2}{h} \mbox{ }\frac{ v_F }{ v_h} }
\end{boxed}
\]
A comparison with standard g-ology with the present chosen model gives the following identifications (Eq.(2.105) of Giamarchi \cite{giamarchi2004quantum}).
\begin{equation*}
\begin{aligned}
&g_{1,\perp} = g_{1,\parallel} = 0
\\&
g_{2,\perp} = g_{2,\parallel} = g_{4,\perp} = g_{4,\parallel} = v_0
\\&
g_{ \rho } = g_{1,\parallel} - g_{2,\parallel} - g_{ 2, \perp} = 0-v_0-v_0 = -2v_0
\\&
g_{ \sigma }= g_{1,\parallel} - g_{2,\parallel} + g_{ 2, \perp}= 0-v_0+v_0 = 0
\\
&
g_{4,\rho} = g_{4,\parallel}+ g_{ 4,\perp} = 2 v_0
\\&
g_{4,\sigma} = g_{4,\parallel} - g_{ 4,\perp} = 0
\\&
y_{ \rho } = g_{ \rho }/( \pi v_F )  = - \frac{2 v_0 }{ \pi v_F }
\\&
y_{ \sigma } = g_{ \sigma } / ( \pi v_F )  = 0
\\&
y_{4,\rho} = g_{4,\rho }/(\pi v_F) = g_{4,\rho }/(\pi v_F) =  2 v_0/(\pi v_F)
\\&
y_{4,\sigma} = g_{4,\sigma }/(\pi v_F) = 0
\end{aligned}
\end{equation*}
\begin{equation*}
\begin{aligned}
u_{ \rho } =& v_F \sqrt{ (1+y_{4,\rho}/2)^2 -(y_{\rho}/2)^2 }\\
 =& v_F \sqrt{ 1+2v_0/(\pi v_F) } \equiv v_h
\end{aligned}
\end{equation*}
\begin{equation*}\small
\begin{aligned}
K_{ \rho } =&  \sqrt{ \frac{1 + y_{4,\rho}/2+y_{\rho}/2}{1 + y_{4,\rho}/2-y_{\rho}/2} }
 = \sqrt{ \frac{1 }{1 +  2v_0/(\pi v_F)} } = \frac{ v_F }{ v_h }
\end{aligned}
\end{equation*}\normalsize
\[
u_{ \sigma }  = v_F \sqrt{ (1+ y_{4,\sigma}/2)^2 - (y_{\sigma}/2)^2 } = v_F
\]
\[
K_{\sigma} = \sqrt{  \frac{1  + y_{4,\sigma}/2 + y_{\sigma}/2 }{1  + y_{4,\sigma}/2 - y_{\sigma}/2  }  } = 1
\]

This gives,

\[
\begin{boxed}
{G  = \frac{ e^2}{h} \mbox{ }\frac{ v_F }{ v_h}  = \frac{ e^2}{h} \mbox{ } K_{\rho}}
\end{boxed}
\]
which is the standard result for a clean quantum wire.

\subsubsection{ The general case: $ |R| > 0 $ and $ v_0 \neq 0 $ }

Again, using the Green function from equation (\ref{SS}) for general value of $|R|$, the current current commutation relation can be calculated as,
\begin{equation*}
\begin{aligned}
<[&j_s(x,\sigma,t),j_s(x',\sigma',t')]> \\
=& - (2 \pi i) \frac{v_F v_h^2 }{ 8\pi^2 v_h } \mbox{   }\partial_{v_h t'} \sum_{  \nu = \pm 1 }\bigg ( \delta ( \nu(x-x') +  v_h(t-t') )
\\&\hspace{1.5cm}- \frac{v_F }{v_h}  \mbox{    }Z_h\mbox{   }
\delta  ( \nu(|x|+|x' |) +  v_h(t-t')  )
\bigg)
\\
&
 - (2 \pi i)
\frac{\sigma\sigma' v_F^2}{ 8\pi^2 } \mbox{   } \partial_{v_Ft'}\sum_{  \nu = \pm 1 }\bigg ( \delta ( \nu(x-x') + v_F(t-t')   )
	\\&\hspace{1.5cm}-|R|^2 \delta  ( \nu(|x|+|x' |) +  v_F(t-t')  )
\bigg)
\end{aligned}
\end{equation*}
where,
\[
\hspace{1 in} Z_h = \frac{ |R|^2 }{    \bigg( 1 - \frac{(v_h-v_F)}{ v_h }
 |R|^2   \bigg) }
\]
Thus,
\begin{equation*}
\begin{aligned}
<j(x,\sigma,t)>  =& ie \sum_{ \sigma^{'} }
\int^{L/2}_{-L/2} dx^{'}\mbox{  }  \int_{-\infty}^{t} dt^{'}\partial_{v_ht^{'}} \mbox{  }\frac{ V_g }{ L}
   (2 \pi i) \frac{v_F  }{ 8\pi^2 } \mbox{   }\\
&\sum_{  \nu = \pm 1 }\bigg ( \theta( -\nu(x-x') -  v_h(t-t') )
\\&- \frac{v_F }{v_h}  \mbox{    }Z_h\mbox{   }
\theta  ( -\nu(|x|+|x' |) - v_h(t-t')  )
\bigg)
\end{aligned}
\end{equation*}
therefore,
\[
<j(x,\sigma,t)>  =
\frac{2 ie }{v_h}
 V_g
   (2 \pi i) \frac{v_F  }{ 8\pi^2 } \mbox{   }\bigg (1
- \frac{v_F }{v_h}  \mbox{    }Z_h
\bigg)
\]
The conductance of a quantum wire without leads but in the presence of barriers and wells is,
\[
G = \frac{ e^2 }{(2\pi)}
    \frac{v_F  }{ v_h  } \mbox{   }\bigg (1
- \frac{v_F }{v_h}  \mbox{    }Z_h
\bigg)
\]
Hence the general formula for the conductance of a quantum wire without leads but with electrons experiencing forward scattering short-range mutual interactions
and in the presence of a finite number of barriers and wells clustered around an origin is (in proper units),
\begin{equation}
\begin{boxed}
{G = \frac{ e^2 }{h}
    \frac{v_F  }{ v_h  } \mbox{   }\bigg (1
- \frac{v_F }{v_h}  \mbox{    }Z_h
\bigg)}
\end{boxed}
\end{equation}
The above general formula agrees with the three well known limiting cases.
\\ \mbox{  } \\
(i) when $ v_h = v_F $, Landauer's formula $ G = \frac{ e^2 }{ h } \mbox{  }|T|^2 $ is recovered.
\\ \mbox{  } \\
(ii) when $ |R| = 0 $, the formula $ G = \frac{ e^2 }{ h }  \mbox{  }K_{\rho} $ is also recovered.
\\ \mbox{  } \\
(iii) when $ |R| = 1 $, $ G = 0 $ regardless of what $ v_h $ is.
\\ \mbox{  }  \\

\subsection{ Conductance from a tunneling experiment }

If the conduction process is envisaged as a tunneling phenomenon as against the usual Kubo formula based approach which involves relating conductance to current-current correlation, a qualitatively different formula for the conductance is obtained.

First observe that the quantity $ |T|^2 $ and $ K_{ \rho  } $ both serve as a ``transmission coefficient" - the former when mutual interactions  are absent but barriers and wells are present and the latter vice versa. Both these may be related to spectral function of the field operator (single particle spectral function) as follows.
\begin{equation*}
\begin{aligned}
&v_F\int^{\infty}_{-\infty}dt\mbox{   }<\{ \psi_{ \nu } (x,\sigma,t) ,  \psi^{\dagger}_{ \nu } (x',\sigma,0)  \}>
 \\&= -(2\pi i)\sum_{ \gamma,\gamma^{'} = \pm 1 } \theta( \gamma x ) \theta( \gamma^{'} x^{'}) g_{ \gamma,\gamma^{'} }(\nu,\nu)
\\&
v_F\int^{\infty}_{-\infty}dt\mbox{   }<\{ \psi_{ \nu } (\nu \frac{L}{2},\sigma,t) ,  \psi^{\dagger}_{ \nu } (-\nu \frac{L}{2},\sigma,0)  \}>
\\&= -(2\pi i)g_{ \nu,-\nu }(\nu,\nu)
\\&
v_F\int^{\infty}_{-\infty}dt\mbox{   }<\{ \psi_{ \nu } (\nu \frac{L}{2},\sigma,t) ,  \psi^{\dagger}_{ \nu } (-\nu \frac{L}{2},\sigma,0)  \}>
 = T
\end{aligned}
\end{equation*}
where $g_{ \gamma,\gamma^{'} }(\nu,\nu)$ are functions of the reflection (R) and the transmission (T) amplitudes of the system and is given explicitly as follows.
\footnotesize
\begin{equation}
\begin{aligned}
\hspace*{-0.2 cm}
\label{gexp}
g_{\gamma_1,\gamma_2} (\nu_1,\nu_2)=\frac{i}{2\pi}& \Big[ \delta_{\nu_1,\nu_2} \delta_{\gamma_1,\gamma_2} \\
&+(T \delta_{\nu_1,\nu_2}+R \delta_{\nu_1,-\nu_2})\delta_{\gamma_1,\nu_1}\delta_{\gamma_2,-\nu_2}\\
&+(T^{*} \delta_{\nu_1,\nu_2}+R^{*} \delta_{\nu_1,-\nu_2})\delta_{\gamma_1,-\nu_1}\delta_{\gamma_2,\nu_2}\Big]
\end{aligned}
\end{equation}
\normalsize
From this point of view, the conductance is related to the magnitude of the above complex number. Choosing it to be proportional to the magnitude of the complex number (rather than the square of the magnitude) allows perfect agreement with the RG equations of Matveev et al. \cite{matveev1993tunneling} as we have seen in the main text ($|T|$ is the magnitude of the transmission amplitude of free fermions plus impurity):\small
\begin{equation}
G = \frac{ e^2 }{h} \mbox{   }|T|\mbox{ }
| v_F\int^{\infty}_{-\infty}dt\mbox{   }<\{ \psi_{ R } ( \frac{L}{2},\sigma,t) ,  \psi^{\dagger}_{ R } (-\frac{L}{2},\sigma,0)  \}>
 |
 \label{TUNNEL}
\end{equation}\normalsize
Note that the above formula is {\bf{not related}} to the square of the dynamical density of states. The dynamical density of states is
equal-space and unequal time Green function. For tunneling, an electron is injected at $ x = - L/2 $
and collected at $ x^{'}  = + L/2 $ as is the case here which is unequal-space unequal-time Green function i.e. the Green function for the electron traversing the impurity.
Technically speaking, the g-ology methods are able to handle only the no barrier case and the half line case properly hence for a weak link they are sometimes forced to surmise that conductance has something to do with dynamical density of states for a half line near the weak link. The present approach is not only different but physically more sensible and compelling. Using the Green function from equation (\ref{OS}),
\footnotesize
\begin{equation*}
\begin{aligned}
\Big\langle T\psi_R(\frac{L}{2},\sigma,t)&\psi_R^{\dagger}(-\frac{L}{2},\sigma,0)\Big\rangle
=\frac{i}{2\pi}\mbox{  }e^{-\frac{1}{2} \log{[L-v_Ft]}}
\\&\times e^{-\frac{1}{2} \log{[L-v_ht]}}e^{-\frac{(v_h-v_F)^2}{8 v_h v_F} \log{\vline \frac{ L^2-(v_ht)^2 }{ L_{ \omega }^2 }\vline }}
\end{aligned}
\end{equation*}
\normalsize
Hence,
\footnotesize
\begin{equation*}
\begin{aligned}
\Big\langle \{\psi_R(\frac{L}{2}&,\sigma,t),\psi_R^{\dagger}(-\frac{L}{2},\sigma,0)\}\Big\rangle
=\frac{i}{2\pi}\mbox{  }e^{-\frac{1}{2} \log{[L-v_F(t-i\epsilon)]}}\\
&\times e^{-\frac{1}{2} \log{[L-v_h(t-i\epsilon)]}}e^{-\frac{(v_h-v_F)^2}{8 v_h v_F} \log{\vline \frac{ L^2-(v_h(t-i\epsilon))^2 }{ L_{ \omega }^2 }\vline }}\\
&\hspace{1 in} - \frac{i}{2\pi}\mbox{  }e^{-\frac{1}{2} \log{[L-v_F(t+i\epsilon)]}}
\\&\times e^{-\frac{1}{2} \log{[L-v_h(t+i\epsilon)]}}e^{-\frac{(v_h-v_F)^2}{8 v_h v_F} \log{\vline \frac{ L^2-(v_h(t+i\epsilon))^2 }{ L_{ \omega }^2 }\vline }}
\end{aligned}
\end{equation*}\normalsize
while integrating over $ t $ the  only regions that contribute are $ L-v_F t \approx 0 $ and $ L - v_h t \approx 0 $. When $ v_h \neq v_F $ these two are different regions. Set $ L - v_F t = y $ then $ L - v_h t =
L - \frac{v_h}{v_F} (L-y) $ and  $ L + v_h t =
L + \frac{v_h}{v_F} (L-y) $. The implication is, integration over $ t $ is now integration over $ y $ and this is important only when $ y $ is close to zero. Next set $ L - v_h t = y^{'} $ then $ L + v_h t = 2L -y^{'} $ and
 $ L - v_F t =
L - \frac{v_F}{v_h} (L-y^{'}) $ and the integrals are important only when $ y^{'} $ is close to zero. This means,
\small
\begin{equation*}
\begin{aligned}
v_F &\int^{\infty}_{-\infty } dt \mbox{   }\Big\langle \{\psi_R(\frac{L}{2},\sigma,t),\psi_R^{\dagger}(-\frac{L}{2},\sigma,0)\}\Big\rangle
\\=&
 \int^{\infty}_{-\infty } dy \mbox{   }\frac{i}{2\pi}\mbox{  }
 \left( e^{-\frac{1}{2} \log{[y+v_Fi\epsilon]}} - e^{-\frac{1}{2} \log{[y-v_Fi\epsilon]}} \right) \mbox{  }\\
&\hspace{1.2cm}e^{-\frac{1}{2} \log{[L (1- \frac{v_h}{v_F}) + \frac{v_h}{v_F} y ]}}
e^{-\frac{(v_h-v_F)^2}{8 v_h v_F} \log{\vline \frac{ L^2- \frac{v^2_h}{v^2_F} (L-y)^2 }{ L_{ \omega }^2 }\vline }}
\\
+& \frac{v_F}{v_h} \int^{\infty}_{-\infty } dy^{'} \mbox{   }\frac{i}{2\pi}\mbox{  }\left( e^{-\frac{1}{2} \log{[y^{'} + v_h i\epsilon ]}} -e^{-\frac{1}{2} \log{[y^{'} - v_h i\epsilon ]}} \right)\\
&\hspace{1.2cm}e^{-\frac{1}{2}\log{[L (1- \frac{v_F}{v_h}) + \frac{v_F}{v_h} y^{'} ]}}
 \mbox{  } e^{-\frac{(v_h-v_F)^2}{8 v_h v_F} \log{\vline \frac{ y^{'}(2L-y^{'}) }{ L_{ \omega }^2 }\vline }}
\end{aligned}
\end{equation*}
\normalsize
Only the dependence on $ L $ is of interest. Write $ y = L \mbox{  }s $ and $ y^{'} = L \mbox{  } s^{'}$. Hence,
\begin{equation*}
\begin{aligned}
v_F \int^{\infty}_{-\infty } dt& \mbox{   }\Big\langle \{\psi_R(\frac{L}{2},\sigma,t),\psi_R^{\dagger}(-\frac{L}{2},\sigma,0)\}\Big\rangle\\	
&\sim e^{-\frac{(v_h-v_F)^2}{8 v_h v_F} \log{\vline \frac{ L^2 }{ L_{ \omega }^2 }\vline }}
\end{aligned}
\end{equation*}
This means the tunneling conductance of a clean (no barrier) quantum wire scales as,
\begin{equation*}
\begin{aligned}
G_{clean} \sim \frac{e^2}{h } &\mbox{  } \mbox{  }
e^{-\frac{(v_h-v_F)^2}{4 v_h v_F} \log{\vline \frac{ L }{ L_{ \omega } }\vline }}
\sim& \left( \frac{ L_{ \omega } }{ L } \right)^{ \frac{1}{4} ( K_{ \rho } + \frac{1}{ K_{ \rho } } - 2 ) }
\end{aligned}
\end{equation*}
\normalsize
where $ L_{ \omega }  = \frac{ v_F }{ k_B T } $ is the length scale associated with temperature
(or frequency since $ k_BT $ is interchangeable with $ \omega $). It says that at low temperatures, the tunneling d.c. conductance of a clean quantum wire with no leads but
  with interactions ($ v_h \neq v_F $) diverges as a power law with exponent $ \frac{1}{4} ( K_{ \rho } + \frac{1}{ K_{ \rho } } - 2 ) > 0 $.
  Fortuitously, the magnitude of this exponent matches with the exponent of the dynamical density of states of a clean wire (no impurity). However when impurities (or a weak link) is present, there is no guarantee that this coincidence will persist.
   For a clean wire there is nothing for a electron to tunnel across so this exercise is pointless. What should be studied is tunneling across a weak link. The general case involves including a finite number finite barriers and wells clustered around the origin. This case is solved elegantly here where a closed formula for the conductance exponents may be obtained
unlike in competing approaches found in the literature where a combination of RG and other approaches are needed that fall well short of providing a closed expression for the exponents. {\it{ More importantly, the present approach is able to provide an analytical interpolation from the weak barrier limit (see above) to the weak link limit to be discussed below - something the competing approaches are incapable of doing without solving complicated RG flow equations, often numerically. }}

In the general case with the barriers and wells, the Green function for points on opposite sides of the origin has a form that is qualitatively different from the form when the points are on the same side of the origin. This is the really striking prediction of this work.

\subsection{ With the impurities }

\noindent Consider the general Green function for $ xx^{'} < 0 $ (equation (\ref{OS})). From that it is possible to conclude
($ W = g_{1,-1}(1,1)\theta(x)\theta(-x')+g_{-1,1}(1,1)\theta(-x)\theta(x') $),
\begin{equation}
\begin{aligned}
<T\psi_R(\frac{L}{2}&,\sigma,t)\psi_R^{\dagger}(-\frac{L}{2},\sigma,0)>
=\frac{v_F+v_h}{2 \sqrt{v_F v_h}}  \mbox{  }g_{1,-1}(1,1)\mbox{  }\\
&e^{(2X+2C)\log{[L]}}\mbox{  }e^{-\frac{1}{2} \log{[L-v_Ft ]}}e^{- \frac{1}{2}\log{[L-v_ht]}}\\
&e^{- (Q-X)
\log{[L^2-(v_ht)^2]}}e^{-C
\log{[-(v_ht)^2]}}
\end{aligned}
\end{equation}
Since $ G \sim | v_F \int^{ \infty }_{-\infty } dt <\{\psi_R(\frac{L}{2},\sigma,t),\psi_R^{\dagger}(-\frac{L}{2},\sigma,0)\}> | $ it is possible to read off the conductance exponent as follows,
\begin{equation}
G \sim \left( \frac{ L }{ L_{ \omega } }\right)^{-2Q }  \mbox{  }  \left( \frac{ L }{ L_{ \omega} }\right)^{ 4X }
\label{GGEN}
\end{equation}
where $ Q=\frac{(v_h-v_F)^2}{8 v_h v_F} $ and
 $ X=\frac{|R|^2(v_h-v_F)(v_h+v_F)}{8  v_h (v_h-|R|^2 (v_h-v_F))} $.
\\
It is easy to see that for a vanishing barrier $ |R| \rightarrow 0 $, the earlier result of the conductance of a clean quantum wire is recovered. The other interesting limit is the weak link limit where $ |R| \rightarrow 1 $. The limiting case of the weak link are two semi-infinite wires.
In this case,
\begin{equation}
G_{weak-link} \sim \left( \frac{ L }{ L_{ \omega} }\right)^{ \frac{ (v_h + v_F)^2-4v^2_F }{ 4 v_h v_F } }
\label{GGEN}
\end{equation}
Hence the d.c. conductance scales as $ G_{weak-link} \sim (k_B T)^{ \frac{ (v_h + v_F)^2-4v^2_F }{ 4 v_h v_F } } $.  This formula is consistent with the
assertions of Kane and Fisher ( C. L. Kane and Matthew P. A. Fisher
Phys. Rev. Lett. {\bf{68}}, 1220 (1992) \cite{kane1992transport}) that show that at low temperatures $ k_B T \rightarrow 0 $
for a fixed $ L $, the conductance vanishes as a power law in the temperature if the interaction between the fermions is repulsive ($ v_h > v_F  > 0 $) and diverges as a power law if the interactions between the fermions is attractive ($ v_F > v_h > 0 $). Their result is applicable to spinless fermions without leads $ G_{weak-link-nospin} \sim (k_B T)^{ \frac{2}{K} - 2 } $ to compare with the result of the present work this exponent has to be halved $  G_{weak-link-with-spin} \sim (k_B T)^{ \frac{1}{K_{ \rho } } - 1 }  $.
This exponent is the same as what we have derived since $ \frac{ (v_h + v_F)^2-4v^2_F }{ 4 v_h v_F } \approx \frac{1}{K_{ \rho } } - 1 $ so long as $ v_h \approx v_F $ (weak interactions).
 In general, the claim of the present work  is that the temperature dependence of the tunneling d.c. conductance of a wire with no leads in the presence of barriers and wells and mutual interaction between particles is,
\[
G \sim (k_B T)^{ \eta} ;\mbox{   }\mbox{  } \mbox{  } \eta = 4X - 2 Q
\]

When $ \eta > 0 $ the conductance vanishes at low temperatures as a power law - characteristic of a weak link. However when $ \eta < 0 $ the conductance diverges at low temperature as a power law - characteristic of a clean quantum wire. This result should not be taken too literally since it is based on the general validity of the surmise in Eq.(\ref{TUNNEL}). This divergence should be taken as an indication of a saturation to a non-zero value.
Of special interest is the situation $ \eta = 0 $ where the  conductance is independent of temperature. This crossover from a conductance that vanishes as a power law at low temperatures to one that diverges as a power law occurs at reflection coefficient
 $ |R|^2 = |R_c|^2 \equiv \frac{v_h (v_h-v_F)}{3 v_F^2+v_h^2} $ which is valid only for repulsive interactions $ v_h > v_F $. For attractive interactions, $ \eta < 0 $ for any $ |R|^2 $ which means
 the conductance always diverges as a power law at low temperatures. This means attractive interactions heal the chain for all reflection coefficients including in the extreme weak link case.
 On the other hand for repulsive interactions, for $ |R| > |R_c| $, $ \eta > 0 $ and the chain is broken (conductance vanishes) at low temperatures. For $ |R| < |R_c| $, $ \eta < 0 $ and even though the interactions are repulsive the chain is healed (conductance diverges).

Note that this section that calculates conductance is based on a serendipitous surmise equation (\ref{TUNNEL}) which equates the tunneling conductance to a certain integral over the one-particle Green function. In hindsight, this surmise works only for temperatures small compared to the bandwidth and for repulsive interactions. Strictly speaking we have to apply a bias and properly calculate the current flowing in a system with bias, impurity, finite-bandwidth interactions and finite temperature. Not surprisingly this is an ambitious project that will lead to a proper formula for the current flowing as a function of the bias and all the other parameters. We expect to recover the RG formulas of Matveev, Yue and Glazman in the limit of weak interactions for a general bandwidth and both attractive and repulsive interactions (not infinite bandwidth repulsive interactions like we have have done in the present manuscript). The main purpose of including this section is just to support the main result namely the Green function of the system. For this the derivation of Eq.(8) of  Matveev, Yue and Glazman as we have been successful in doing in the main text is already sufficient.

\section*{Funding}
A part of this work was done with financial support from Department of Science and Technology, Govt. of India DST/SERC: SR/S2/CMP/46 2009.\\

\bibliographystyle{apsrev4-1}
\bibliography{ref}
\normalsize

\end{document}